%% file: belle-conf-0472.tex
\documentclass[aps,prl,preprint,tightenlines,superscriptaddress,showpacs,byrevtex]{revtex4}

\usepackage{epsfig, amssymb}

\begin{document}
\advance\hoffset by  -4mm

\newcommand{\de}{\Delta E}
\newcommand{\mbc}{M_{\rm bc}}
\newcommand{\mds}{M(D_s)}
\newcommand{\bb}{B{\bar B}}
\newcommand{\qq}{q{\bar q}}
\newcommand{\kstar}{K^{*0}}
\newcommand{\kpi}{K^+\pi^-}
\newcommand{\kk}{K^+K^-}
\newcommand{\kkpi}{K^+K^-\pi^+}
\newcommand{\phipi}{\phi\pi^+}
\newcommand{\phik}{\phi K^-}
\newcommand{\kstark}{{\bar K^{*0}}K^+}
\newcommand{\ksk}{K_S^0 K^+}
\newcommand{\kspi}{K_S^0 \pi^+}
\newcommand{\kstarpip}{{\bar K^{*0}}\pi^+}
\newcommand{\kpipi}{K^-\pi^+\pi^+}
\newcommand{\bdsk}{{\bar B^0}\to D_s^+K^-}
\newcommand{\bdspi}{B^0\to D_s^+\pi^-}
\newcommand{\bdsstarpi}{B^0\to D^{*+}_s\pi^-}
\newcommand{\bdsh}{\bar{B^0}\to D_s^+ h^-}
\newcommand{\bdsstarh}{{\bar B^0}\to D^{*+}_s h^-}
\newcommand{\dpkpipi}{D^+\to\kpipi}
\newcommand{\dsphipi}{D_s^+\to\phipi}
\newcommand{\dskstark}{D_s^+\to\kstark}
\newcommand{\dsksk}{D_s^+\to\ksk}
\newcommand{\bdppi}{{\bar B^0}\to D^+\pi^-}
\newcommand{\br}{{\cal B}}
\newcommand{\tim}{\times 10^{-5}}
\newcommand{\brdsk}{2.93\pm 0.55\pm 0.79}
\newcommand{\brdspi}{1.94\pm 0.47\pm 0.52}
\newcommand{\brdskphipi}{(10.5\pm 2.0\pm 1.0)\times 10^{-7}}
\newcommand{\brdspiphipi}{(7.0\pm 1.7\pm 0.7)\times 10^{-7}}

\preprint{\vbox{ \hbox{   }
    \hbox{BELLE-CONF-0472}
}}

\title{\Large \rm Observation of $\bdspi$}
\input {author-conf2004.tex}
\noaffiliation
 
\begin{abstract}
We report the first observation of $\bdspi$ and an improved
measurement of $\bdsk$ based on $274\times 10^6$ $\bb$ events
collected with the Belle detector at KEKB.
We measure the branching fractions 
$\br(\bdsk)=(\brdsk)\tim$ and $\br(\bdspi)=(\brdspi)\tim$.
\end{abstract}
\maketitle

The unitarity of the Cabbibo-Kobayashi-Maskawa (CKM) matrix~\cite{ckm} 
is a crucial component of the Standard Model, one that is
currently being tested in multiple aspects through $B$ meson decays
at the $B$-factories.
The decays $\bdspi$ are expected to proceed dominantly through a
spectator process involving the $b\to u$ transition and are thus 
considered a promising mode for the precise determination of 
$|V_{ub}|$~\cite{vub}.
The decay $\bdsk$ is not directly accessible through a spectator 
process, requiring either final state interactions or a non-spectator 
$W$-exchange~\cite{dsk}.
The decay $\bdsk$ was observed by Belle~\cite{belle_dsh} and confirmed by
BaBar~\cite{babar_dsh}. For $\bdspi$, both groups reported 
evidence~\cite{belle_dsh,babar_dsh}.

In this Letter we report improved measurements of $\bdsk$ and $\bdspi$
decays with the Belle detector~\cite{NIM} at the KEKB
asymmetric energy $e^+e^-$ collider~\cite{KEKB}.
The results are based on a data sample, collected at the 
center-of-mass (CM) energy of the $\Upsilon(4S)$ resonance, which
contains $274\times 10^6$ produced $B\bar B$ pairs.

The Belle detector is a large-solid-angle magnetic
spectrometer that consists of a silicon vertex detector (SVD),
a 50-layer central drift chamber (CDC), an array of
aerogel threshold \v{C}erenkov counters (ACC),
a barrel-like arrangement of time-of-flight
scintillation counters (TOF), and an electromagnetic calorimeter (ECL)
comprised of CsI(Tl) crystals located inside
a superconducting solenoid coil that provides a 1.5~T
magnetic field.  An iron flux-return located outside of
the coil is instrumented to detect $K_L^0$ mesons and to identify
muons (KLM).  The detector is described in detail elsewhere~\cite{NIM}.
Two different inner detector configurations were used. For the first sample
of 152 million $B\bar{B}$ pairs, a 2.0 cm radius beampipe
and a 3-layer silicon vertex detector were used;
for the latter 122 million $B\bar{B}$ pairs,
a 1.5 cm radius beampipe, a 4-layer silicon detector
and a small-cell inner drift chamber were used\cite{Ushiroda}.

Charged tracks are selected with requirements based on the
average hit residual and impact parameter relative to the
interaction point (IP). We also require that the transverse momentum of
the tracks be greater than 0.1 GeV$/c$ in order to reduce the low 
momentum combinatorial background.
For charged particle identification (PID) the combined information
from specific ionization in the central drift chamber ($dE/dx$), 
time-of-flight scintillation counters (TOF) and aerogel \v{C}erenkov 
counters (ACC) is used.
At large momenta ($>2.5$~GeV$/c$) only the ACC and $dE/dx$ are used.
Charged kaons are selected with PID criteria that have
an efficiency of 88\%, a pion misidentification probability of 8\%,
and negligible contamination from protons.
The criteria for charged pions have an efficiency of 89\% and 
a kaon misidentification probability of 9\%.
All tracks that are positively identified as electrons are rejected.

Neutral kaons are reconstructed via the decay $K_S^0\to\pi^+\pi^-$.
The two-pion invariant mass is required to be within 6~MeV$/c^2$ 
($\sim 2.5\sigma$) of the nominal $K^0$ mass, and the displacement of 
the $\pi^+\pi^-$ vertex from the IP in the transverse 
$r$-$\phi$ plane is required to be between 0.1~cm and 20~cm. 
The directions in the $r$-$\phi$ projection of the $K_S^0$ candidate's 
flight path and momentum are required to agree within 0.2 radians.

We reconstruct $D_s^+$ mesons in the channels
$D_s^+\to \phipi$, $\kstark$, and $\ksk$ (inclusion of charge
conjugate states is implicit throughout this report).
$\phi$ ($\kstar$) mesons are formed from the $\kk$ ($\kpi$ ) pairs with
invariant mass within 10~MeV$/c^2$ (50~MeV$/c^2$) of the nominal
$\phi$ ($\kstar$) mass.
We select $D_s^+$ mesons in a wide ($\pm 0.5$~GeV$/c^2$) window,
for subsequent studies; the $\mds$
signal region is defined to be within 12~MeV$/c^2$ ($\sim 2.5\sigma$)
of the nominal $D_s^+$ mass. $D_s^+$ candidates are combined with a 
charged kaon or pion to form a $B$ meson.
Candidate events are identified by their CM
energy difference, \mbox{$\de=(\sum_iE_i)-E_b$}, and the
beam constrained mass, $\mbc=\sqrt{E_b^2-(\sum_i\vec{p}_i)^2}$, where
$E_b = \sqrt{s}/2$ is the beam energy and 
$\vec{p}_i$ and $E_i$ are the momenta and
energies of the decay products of the $B$ meson in the CM frame.
We select events with $\mbc>5.2$~GeV$/c^2$
and $|\de|<0.2$~GeV and define the $B$ signal region to be
$5.272$~GeV$/c^2<\mbc<5.288$~GeV$/c^2$ and $|\de|<0.03$~GeV.
The $\mbc$ sideband is defined as $5.20$~GeV$/c^2<\mbc<5.26$~GeV$/c^2$.
We use a Monte Carlo (MC) simulation to determine the efficiency~\cite{GEANT}.

To suppress the large combinatorial background that is dominated by 
the two-jet-like $e^+e^-\to\qq$ ($q=u,d,s$ and $c$ quarks) 
continuum 
process, variables that characterize the event topology are used. 
We require $|\cos\theta_{\rm thr}|<0.80$, where $\theta_{\rm thr}$ is 
the angle between the thrust axis of the $B$ candidate and that of the 
rest of the event.  This requirement eliminates 77\% of the continuum 
background and retains 78\% of the signal events. We also define a 
Fisher discriminant, ${\cal F}$, 
which is based on the production angle of 
the $B$ candidate, the angle of the $B$ candidate thrust axis with 
respect to the beam axis, and nine parameters that characterize the 
momentum flow in the event relative to the $B$ candidate thrust axis 
in the CM frame~\cite{VCal}. We impose a requirement on ${\cal{F}}$ 
that rejects 50\% of the remaining continuum background and 
retains 92\% of the signal.

We also consider possible backgrounds from $\qq$ events containing
real $D_s^+$ mesons.  These events peak in the $\mds$ spectra
but not in the $\de$ and $\mbc$ distributions. 
Thus it can be eliminated by fitting the $\de$ distribution.

Other $B$ decays, such as $\bdppi$, $\dpkpipi$, with one pion 
misidentified as a kaon, require particular attention because they 
have large branching fractions and can peak in 
the $\mbc$ signal region. 
The reconstructed invariant mass spectra for these events
overlap with the signal $D_s^+$ mass region, while their $\de$ 
distribution is shifted by about 50~MeV$/c^2$.
To suppress this background, we exclude event candidates 
that are consistent with the $\dpkpipi$ mass hypothesis within 
15~MeV$/c^2$ ($\sim 3\sigma$) when the two same-sign particles
are considered to be pions, independently of their PID information.
For the $\dsksk$ mode there is a similar background from $\bdppi$, 
$D^+\to\kspi$. In this case we
exclude candidates consistent within 20~MeV$/c^2$ ($\sim 3\sigma$)
with the $D^+\to\kspi$ hypothesis. 

Possible backgrounds from  $B$ decays via $b\to c$ transitions 
($B\to D_s D X$) are also considered.
The $D_s$ mesons from these decays have a lower momentum and 
are kinematically separated from the signal.
We analyzed a large MC sample of generic $\bb$ events 
and found no peaking backgrounds.

Another potential $\bb$ background is charmless 
${\bar B^0}\to K^-K^+K^-\pi^+ (K_S^0\kk)$. Such events peak in the
$\de$ and $\mbc$ spectra, but not in the $\mds$ distributions. 
They tend to be dominated by quasi-two-body decay 
channels such as $\phi {\bar K^{(*)0}}$~\cite{PDG}.
To reduce this background, we reject events with low 
($<2$~GeV$/c^2$) two particle invariant masses: $M_{K^-\pi^+}$ and 
$M_{\phik}$ for the $\dsphipi$ channel, 
$M_{\kk}$ and $M_{{\bar K^{*0}}K^-}$ for $\dskstark$, 
and $M_{\kk}$ and $M_{K_S^0 K^-}$ for $\dsksk$.
The remaining background from these sources, if any, is excluded by 
fitting the $\mds$ distribution.

\begin{figure}
  \includegraphics[width=0.7\textwidth] {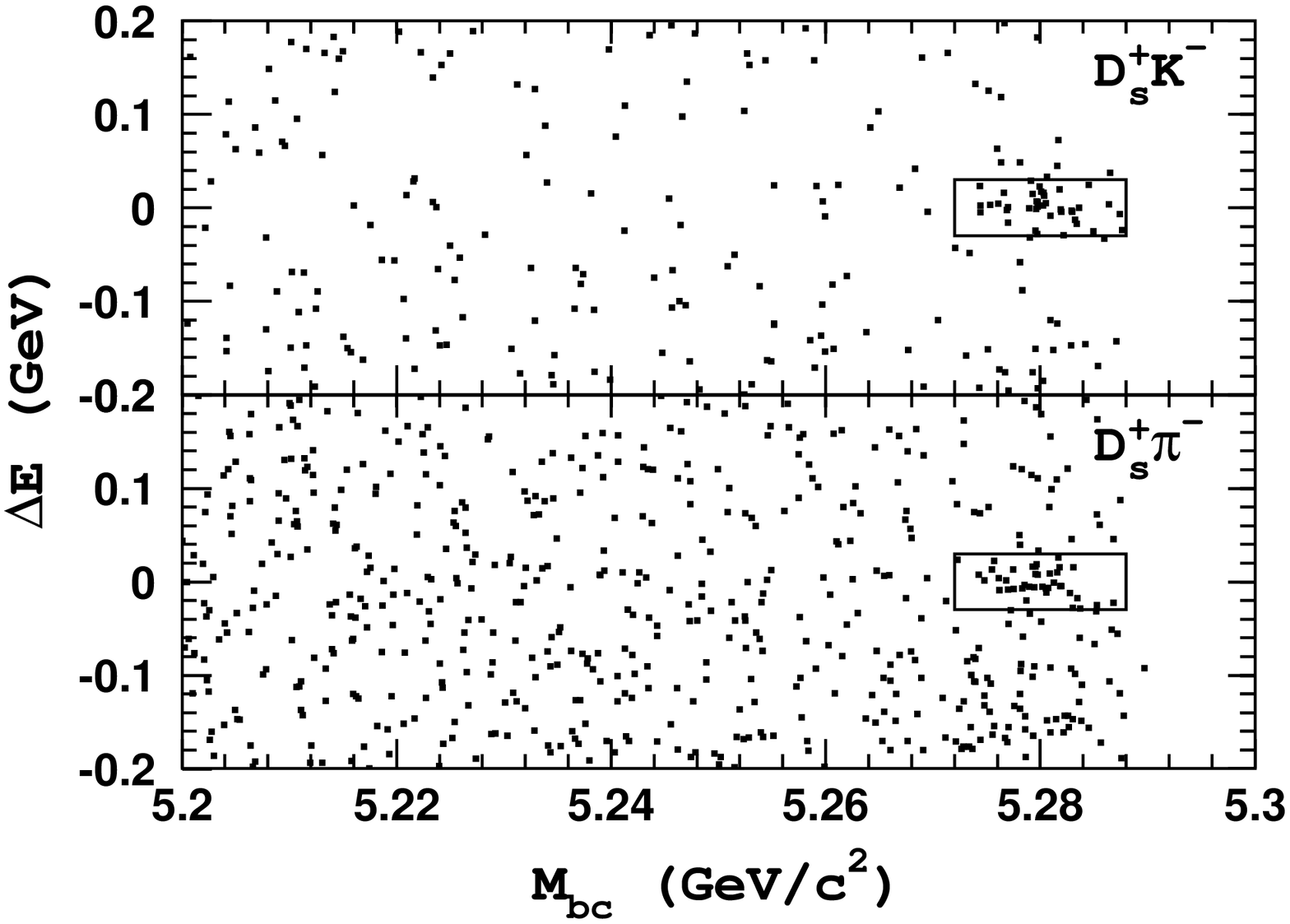} 
  \caption{$\de$ versus $\mbc$ scatter plot for the $\bdsk$ (top) and
    $\bdspi$ (bottom) candidates in the $\mds$ signal region.
  The points represent the
  experimental data and the boxes show the $B$ meson signal region.}
  \label{dsh_mbcde}
\end{figure}

\begin{figure}
  \includegraphics[width=0.7\textwidth] {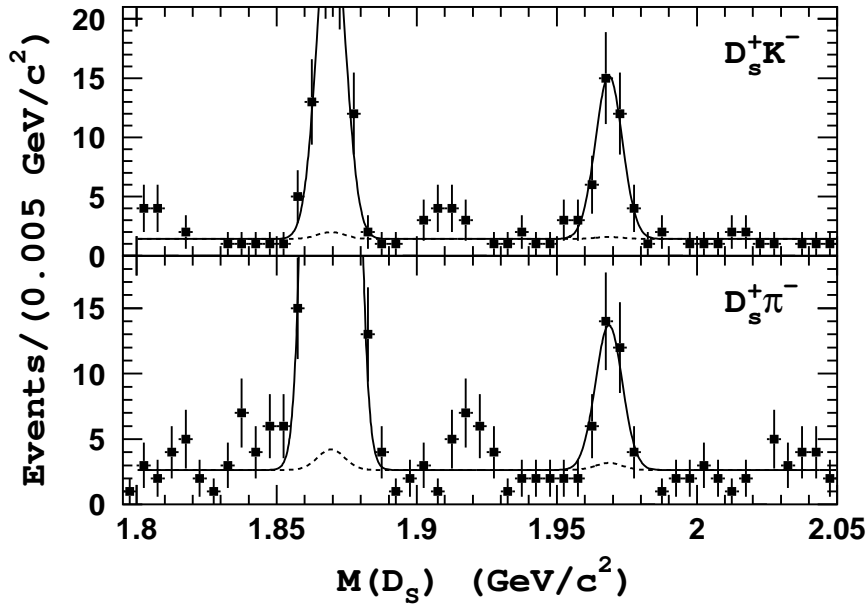} 
  \caption{$\mds$ spectra for $\bdsk$ (top) and
    $\bdspi$ (bottom) in the $B$ signal region. 
    The points with errors represent 
    experimental data and the curves display the results of 
    the simultaneous fit described in the text.
    The small feature near $1.92 \ {\rm GeV}/c^2$ 
    arises due to favoured $D^+$ decays with pions misidentified as kaons.
  }
  \label{dsh_mds}
\end{figure}

The scatter plots in  $\de$ and $\mbc$  for the $\bdsk$ and $\bdspi$ 
candidates in the $\mds$ signal region are shown in
Fig.~\ref{dsh_mbcde}; a significant enhancement in the 
$B$ signal region is observed.
Figure~\ref{dsh_mds} shows the $\mds$ spectra for selected 
$\bdsk$ and $\bdspi$ candidates in the $B$ signal region.
In addition to clear signals at the $D_s^+$ mass in
Fig.~\ref{dsh_mds}, we  observe peaks at the $D^+$ mass, 
corresponding to $\bdppi$ and ${\bar B^0}\to D^+ K^-$, 
$D^+\to\phipi, \kstark, \ksk$. 

Our studies have shown that the backgrounds may peak in the signal
region of $\mds$ or of $\de$ (and $\mbc$) but not in both
simultaneously. To extract our signal, we therefore perform a binned
maximum likelihood fit to the two-dimensional distribution of data in
$\mds$ and $\de$, separating the backgrounds from the signal
component, which peaks in both.
We select events from $\mbc$ signal region for this fit.
For each of the three $D_s^+$ decay channels the $\de$ range,
$-0.1$~GeV$<\de<0.2$~GeV, is divided into 30 bins and the $\mds$
range, $1.5$~GeV$/c^2<\mds<2.5$~GeV$/c^2$, into 200 bins.
All bins in all $D_s^+$ submodes are fitted simultaneously
to a sum of signal and background shapes.
The $D_s^+$ signal is described by a two-dimensional Gaussian, with
widths in both dimensions obtained and fixed using reconstructed
signals in the data from $\bdppi (D^+\to K^-\pi^+\pi^+, K_S \pi^+)$.
The signal amplitude is constrained to correspond to the same
branching fraction $\br(\bdsh)$ for all three $D_s^+$ submodes.
The fit also includes an additional two-dimensional Gaussian 
for ${\bar B^0}\to D^+h^-$ decays.

\begin{table*}
\caption{Results on the signal yields and branching fractions.
Decay channel, signal yield from two dimensional $\mds$ - $\de$,
one dimensional $\mds$ and $\de$ fits, detection efficiency
branching fraction and statistical significance.
The efficiencies include intermediate branching fractions.
}
\footnotesize
\medskip
\label{defit}
  \begin{tabular*}{\textwidth}{l@{\extracolsep{\fill}}cccccc}\hline\hline
 Mode  & $\mds$ - $\de$ & $\mds$ & $\de$  
& $\epsilon$, $10^{-3}$ &
${\cal B}$ $(10^{-5})$ & 
Significance\\\hline
$\bdsk$, $\dsphipi$ & $18.2^{+5.0}_{-4.3}$ &
        $18.2^{+4.7}_{-4.0}$ & $18.1^{+4.9}_{-4.3}$ &
        $2.05$ & $3.25^{+0.89}_{-0.77}\pm 0.88$ & $6.2\sigma$\\

$\bdsk$, $\dskstark$ & $13.1^{+4.5}_{-3.8}$ &
        $11.2^{+4.2}_{-3.6}$ & $13.9^{+4.5}_{-3.9}$ &
        $1.49$ & $3.21^{+1.10}_{-0.93}\pm 0.87$ & $4.6\sigma$\\

$\bdsk$, $\dsksk$ &  $3.7^{+2.8}_{-2.2}$ &
        $3.6^{+2.8}_{-2.1}$ & $5.0^{+3.0}_{-2.3}$ &
        $0.86$ & $1.57^{+1.20}_{-0.94}\pm 0.42$ & $1.8\sigma$\\
\hline
$\bdsk$, simultaneous fit & $36.3^{+7.2}_{-6.6}$ &
        $32.9^{+6.6}_{-5.9}$ & $36.7^{+7.1}_{-6.4}$ & $4.29$ &
        $\brdsk$ & $7.9\sigma$\\\hline
$\bdspi$, $\dsphipi$ & $12.7^{+4.4}_{-3.7}$ &
        $14.4^{+4.2}_{-3.9}$ & $11.7^{+4.3}_{-3.7}$ & 
        $2.29$ & $2.03^{+0.70}_{-0.59}\pm 0.55$ & $4.5\sigma$\\

$\bdspi$, $\dskstark$ & $7.0^{+4.3}_{-3.6}$ &
        $5.3^{+3.9}_{-3.2}$ & $9.3^{+4.5}_{-3.9}$ & 
        $1.65$ & $1.54^{+0.95}_{-0.79}\pm 0.42$ & $2.1\sigma$\\

$\bdspi$, $\dsksk$ & $5.8^{+3.4}_{-2.7}$ &
        $6.1^{+3.2}_{-2.6}$ & $6.1^{+3.5}_{-2.8}$ &
        $0.89$ & $2.38^{+1.40}_{-1.12}\pm 0.64$ & $2.4\sigma$\\
\hline
$\bdspi$, simultaneous fit & $25.7^{+6.5}_{-6.0}$ &
        $26.3^{+6.4}_{-5.7}$ & $27.3^{+6.9}_{-6.3}$ & $4.83$ &
        $\brdspi$ & $5.5\sigma$\\\hline\hline
  \end{tabular*}
\end{table*}

\begin{figure}
  \includegraphics[width=0.7\textwidth] {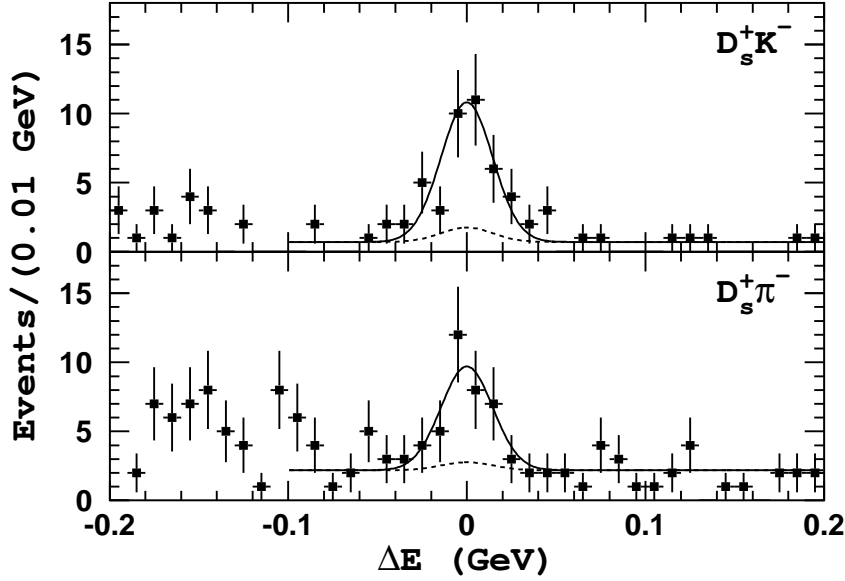}
  \caption{$\de$ spectra for $\bdsk$ (top) and $\bdspi$ (bottom)
    in the $B$ signal region. The points with errors are 
    experimental data and the curves are the results of 
    the simultaneous fit described in the text.}
  \label{dsh_mdsde}
\end{figure}

The background includes three components: combinatorial (linear in 
$\mds$ and $\de$), $\qq$ events that peak in $\mds$ and are flat in 
$\de$, and $B$ decays that peak in $\de$ and are flat in $\mds$.
The levels of the three  components are allowed to vary 
independently in the three reconstructed $D_s^+$ modes.
The fit results are given in Table~\ref{defit}.
The statistical significance quoted in Table~\ref{defit} is defined as 
$\sqrt{-2\ln ({\cal L}_0/{\cal L}_{max})}$, where ${\cal L}_{max}$ and 
${\cal L}_{0}$ denote the maximum likelihood with the fitted signal 
yield and with the signal yield fixed to zero, respectively.
The results of one-dimensional fits to the $\mds$ and $\de$ 
distributions are also shown in Table~\ref{defit} for comparison.
Figures~\ref{dsh_mds} and \ref{dsh_mdsde} show the $\mds$ and $\de$
projections for events from the signal region. The sum of the fitted 
signal plus background is shown by the solid lines while the
background shape including the peaking background is indicated by 
dashed lines. The peaking background is found to be 
$3.6\pm 0.6$ and $3.4\pm 1.3$ events for $\bdsk$ and $\bdspi$, respectively. 

$\bdsstarh$ final states,  where the low energy photon from the 
$D^*_s\to D_s\gamma$ decay is missed,
can populate the $\bdsh$ signal region. 
These  would produce a long tail on the negative side of the $\de$ 
distribution.  
In theoretical models based on factorization, 
the $\bdsstarpi$ and $\bdspi$ decay widths are predicted to be 
approximately equal; there are, however, no 
corresponding predictions  for ${\bar B^0}\to D^{(*)+}_s K^-$ decays.
To study the sensitivity of the measured branching fraction to a 
possible $\bdsstarh$ contribution, we perform a fit with an additional 
$\bdsstarh$ component included, where the signal shape is fixed from the 
MC and the branching fraction is left as a free parameter. The resulting
2\% difference in the $\bdspi$ event yield (compared to the results 
presented in Table~\ref{defit}) is added to the systematic
uncertainty; the change in the $\bdsk$ yield is less than 1\%.
We also check for crossfeed between $\bdsk$ and $\bdspi$
due to kaon/pion misidentification.
To study this we include the crossfeed contributions in the
simultaneous fit, with shapes 
fixed from the MC and misidentification
probabilities obtained from data; 
the uncertainty due to this effect
is found to be  negligible ($\lesssim 1\%$).

As a check, we apply the same procedure to 
$\bdppi$ and ${\bar B^0}\to D^+ K^-$, $D^+\to\phipi, \kstark, \ksk$, 
and obtain 
$\br(\bdppi)=(2.8\pm 0.1)\times 10^{-3}$ and 
$\br({\bar B^0}\to D^+ K^-)=(2.4\pm 0.4)\times 10^{-4}$~\cite{check},
which agree well with the PDG values~\cite{PDG}.

The following sources of systematic error are found to be 
the most significant: tracking efficiency (1-2\% per track), charged 
hadron identification efficiency (1\% per particle), $K^0_S$
reconstruction efficiency (6\%),
signal-shape parameterization (5\%) and MC statistics (2\%).
The tracking efficiency error is estimated using
$\eta$ decays to $\gamma\gamma$ and $\pi^+\pi^-\pi^0$.
The $K/\pi$ identification uncertainty is determined 
from $D^{*+}\to D^0\pi^+$, $D^0\to K^-\pi^+$ decays.
We assume equal production of $B^+B^-$ and $B^0\bar B^0$ pairs but do
not include an additional error from this
assumption.
The uncertainty in the $D_s^+$ meson branching fractions, which is 
dominated by the 25\% error on $\br(\dsphipi)$, is also taken
into account.
The overall systematic uncertainty is 27\%.

In summary, we report the first observation of $\bdspi$ and an
improved measurement of $\bdsk$. We find $\br(\bdsk)=(\brdsk)\tim$ and
$\br(\bdspi)=(\brdspi)\tim$.
Since the dominant systematic uncertainty in both  measurements is due to
the branching fraction for $\dsphipi$, $\br_{\phi\pi}$, we also report
$\br(\bdsk)\times\br_{\phi\pi}=\brdskphipi$ and
$\br(\bdspi)\times\br_{\phi\pi}=\brdspiphipi$.
These results are consistent with previous
results~\cite{belle_dsh, babar_dsh} and have better accuracy.
They supersede results from~\cite{belle_dsh}.

We thank the KEKB group for the excellent operation of the
accelerator, the KEK Cryogenics group for the efficient
operation of the solenoid, and the KEK computer group and
the National Institute of Informatics for valuable computing
and Super-SINET network support. We acknowledge support from
the Ministry of Education, Culture, Sports, Science, and
Technology of Japan and the Japan Society for the Promotion
of Science; the Australian Research Council and the
Australian Department of Education, Science and Training;
the National Science Foundation of China under contract
No.~10175071; the Department of Science and Technology of
India; the BK21 program of the Ministry of Education of
Korea and the CHEP SRC program of the Korea Science and
Engineering Foundation; the Polish State Committee for
Scientific Research under contract No.~2P03B 01324; the
Ministry of Science and Technology of the Russian
Federation; the Ministry of Education, Science and Sport of
the Republic of Slovenia;  the Swiss National Science Foundation; 
the National Science Council and
the Ministry of Education of Taiwan; and the U.S.\
Department of Energy.

\end{document}

%% file: author-conf2004.tex
\affiliation{Aomori University, Aomori}
\affiliation{Budker Institute of Nuclear Physics, Novosibirsk}
\affiliation{Chiba University, Chiba}
\affiliation{Chonnam National University, Kwangju}
\affiliation{Chuo University, Tokyo}
\affiliation{University of Cincinnati, Cincinnati, Ohio 45221}
\affiliation{University of Frankfurt, Frankfurt}
\affiliation{Gyeongsang National University, Chinju}
\affiliation{University of Hawaii, Honolulu, Hawaii 96822}
\affiliation{High Energy Accelerator Research Organization (KEK), Tsukuba}
\affiliation{Hiroshima Institute of Technology, Hiroshima}
\affiliation{Institute of High Energy Physics, Chinese Academy of Sciences, Beijing}
\affiliation{Institute of High Energy Physics, Vienna}
\affiliation{Institute for Theoretical and Experimental Physics, Moscow}
\affiliation{J. Stefan Institute, Ljubljana}
\affiliation{Kanagawa University, Yokohama}
\affiliation{Korea University, Seoul}
\affiliation{Kyoto University, Kyoto}
\affiliation{Kyungpook National University, Taegu}
\affiliation{Swiss Federal Institute of Technology of Lausanne, EPFL, Lausanne}
\affiliation{University of Ljubljana, Ljubljana}
\affiliation{University of Maribor, Maribor}
\affiliation{University of Melbourne, Victoria}
\affiliation{Nagoya University, Nagoya}
\affiliation{Nara Women's University, Nara}
\affiliation{National Central University, Chung-li}
\affiliation{National Kaohsiung Normal University, Kaohsiung}
\affiliation{National United University, Miao Li}
\affiliation{Department of Physics, National Taiwan University, Taipei}
\affiliation{H. Niewodniczanski Institute of Nuclear Physics, Krakow}
\affiliation{Nihon Dental College, Niigata}
\affiliation{Niigata University, Niigata}
\affiliation{Osaka City University, Osaka}
\affiliation{Osaka University, Osaka}
\affiliation{Panjab University, Chandigarh}
\affiliation{Peking University, Beijing}
\affiliation{Princeton University, Princeton, New Jersey 08545}
\affiliation{RIKEN BNL Research Center, Upton, New York 11973}
\affiliation{Saga University, Saga}
\affiliation{University of Science and Technology of China, Hefei}
\affiliation{Seoul National University, Seoul}
\affiliation{Sungkyunkwan University, Suwon}
\affiliation{University of Sydney, Sydney NSW}
\affiliation{Tata Institute of Fundamental Research, Bombay}
\affiliation{Toho University, Funabashi}
\affiliation{Tohoku Gakuin University, Tagajo}
\affiliation{Tohoku University, Sendai}
\affiliation{Department of Physics, University of Tokyo, Tokyo}
\affiliation{Tokyo Institute of Technology, Tokyo}
\affiliation{Tokyo Metropolitan University, Tokyo}
\affiliation{Tokyo University of Agriculture and Technology, Tokyo}
\affiliation{Toyama National College of Maritime Technology, Toyama}
\affiliation{University of Tsukuba, Tsukuba}
\affiliation{Utkal University, Bhubaneswer}
\affiliation{Virginia Polytechnic Institute and State University, Blacksburg, Virginia 24061}
\affiliation{Yonsei University, Seoul}
  \author{K.~Abe}\affiliation{High Energy Accelerator Research Organization (KEK), Tsukuba} 
  \author{K.~Abe}\affiliation{Tohoku Gakuin University, Tagajo} 
  \author{N.~Abe}\affiliation{Tokyo Institute of Technology, Tokyo} 
  \author{I.~Adachi}\affiliation{High Energy Accelerator Research Organization (KEK), Tsukuba} 
  \author{H.~Aihara}\affiliation{Department of Physics, University of Tokyo, Tokyo} 
  \author{M.~Akatsu}\affiliation{Nagoya University, Nagoya} 
  \author{Y.~Asano}\affiliation{University of Tsukuba, Tsukuba} 
  \author{T.~Aso}\affiliation{Toyama National College of Maritime Technology, Toyama} 
  \author{V.~Aulchenko}\affiliation{Budker Institute of Nuclear Physics, Novosibirsk} 
  \author{T.~Aushev}\affiliation{Institute for Theoretical and Experimental Physics, Moscow} 
% ...
\collaboration{The Belle Collaboration}

%% file: belle-conf-0472.bbl
\begin{thebibliography}{99}
\bibitem{check}{These results should not be considered as a measurements
  of the corresponding branching fractions since we do not perform
  analysis of the systematic errors.}
\end{thebibliography}
